\documentclass[]{spie}  %>>> use for US letter paper
\usepackage{amsmath,amsfonts,amssymb}
\usepackage{graphicx}
\usepackage[colorlinks=true, allcolors=blue]{hyperref}
\usepackage{caption}
\usepackage{xspace}
\usepackage{bm}
\usepackage{adjustbox}
\usepackage{tabularx}
\usepackage{color}
\usepackage{algorithm}
\usepackage{algpseudocode}
\usepackage{caption}
\usepackage{subcaption}
\usepackage{listings}
\usepackage{nccmath}
\usepackage{bbm}
\usepackage{multirow}
\usepackage{longtable}
\usepackage{authblk}
\definecolor{seancolor}{rgb}{0.1,0.1,0.8}

\definecolor{francolor}{rgb}{0.8,0.1,0.1}

\definecolor{tonycolor}{rgb}{0.8,0.8,0.1}

\definecolor{varuncolor}{rgb}{0.1,0.8,0.1}

\makeatletter
\DeclareRobustCommand\onedot{\futurelet\@let@token\@onedot} % Add a period to the end of an abbreviation unless there's one already, then \xspace.
\def\@onedot{\ifx\@let@token.\else.\null\fi\xspace}

\def\eg{\emph{e.g}\onedot}

\def\etc{\emph{etc}\onedot}

\makeatother

\title{ST-FL: Style Transfer Preprocessing in Federated Learning for COVID-19 Segmentation}

\author[1,*]{Antonios Georgiadis}
\author[1,2,*]{Varun Babbar}
\author[1]{Fran Silavong}
\author[1]{Sean Moran}
\author[1]{Rob Otter}
\affil[1]{CTO Applied Research: JP Morgan Chase}
\affil[2]{University of Cambridge}
\affil[*]{These authors contributed equally to this work.}
\begin{document} 
\maketitle

\begin{abstract}
Chest Computational Tomography (CT) scans present low cost, speed and objectivity for COVID-19 diagnosis and deep learning methods have shown great promise in assisting the analysis and interpretation of these images. Most hospitals or countries can train their own models using in-house data, however empirical evidence shows that those models perform poorly when tested on new unseen cases, surfacing the need for coordinated global collaboration. Due to privacy regulations, medical data sharing between hospitals and nations is extremely difficult. We propose a GAN-augmented federated learning model, dubbed ST-FL (Style Transfer Federated Learning), for COVID-19 image segmentation. Federated learning (FL) permits a centralised model to be learned in a secure manner from heterogeneous datasets located in disparate private data silos. We demonstrate that the widely varying data quality on FL client nodes leads to a sub-optimal centralised FL model for COVID-19 chest CT image segmentation. ST-FL is a novel FL framework that is robust in the face of highly variable data quality at client nodes. The robustness is achieved by a denoising CycleGAN model at each client of the federation that maps arbitrary quality images into the same target quality, counteracting the severe data variability evident in real-world FL use-cases. Each client is provided with the target style, which is the same for all clients, and trains their own denoiser. Our qualitative and quantitative results suggest that this FL model performs comparably to, and in some cases better than, a model that has centralised access to all the training data.
\end{abstract}

\keywords{Style transfer, federated learning, COVID-19 segmentation}
% Tony: Check if this list needs to be from a pre-determined list.

%%% SUMMARY
\section{DECLARATION}
We hereby attest that this work has not been submitted for publication or presentation in any other conference. 
\section{SUMMARY}
\label{sec:summary}
We developed a noise-agnostic flavour of Federated Learning by utilizing style transfer prepossessing prior to the federation - the ultimate goal is to achieve better results in COVID-19 segmentation, when the models are trained with FL. The target style is shared in the form of a dataset (\eg 20-50 images), and each client trains its own CycleGAN transformation which maps its local data to the common style. The benefit of this approach is that it reduces any image noise in the CT scans prior to sending information via the federation. We used 2 publicly available COVID-19 segmentation datasets, in addition to artificial noise patterns, and demonstrate statistically significant lesion segmentation improvement ranging between 5\%-40\%, depending on the noise pattern.

%%% INTRODUCTION

\section{INTRODUCTION}
\label{sec:intro}  % \label{} allows reference to this section
Due to stringent privacy laws, sharing of confidential data between institutions and countries is fraught with difficulties, and is generally considered impossible. \emph{Federated Learning} provides a solution to this data sharing dilemma, allowing globally distributed data to remain private while still permitting a centralised neural network model to be learnt using information from all of these images existing across institution and country boundaries. Federated learning solves the problem of how to learn a single model based on data that is locked away in data silos without revealing per-client private data to other clients or the central server. The client and the aggregator share the same neural network architecture. Clients train on their local data and send the gradient updates to the aggregator, these gradient updates are combined by the aggregator potentially in a cryptographically secure manner \cite{Bonawitz17}, the central model weights are updated with the aggregared gradients, and the resulting weights are distributed to the clients at the same time.

Prior research has explored the benefits of federated learning for leveraging disparate datasets for the purpose of COVID-19 chest CT scan segmentation~\cite{yang2021federated}. However, there is no previous research that accounts for the differing factors of variation of CT images that are distributed across client nodes. In practice CT images arising from different generations of CT machine can differ vastly across many factors of variation, for example brightness, detail and noise level, in addition to factors such as using a contrast-enhancing agent prior to the scan (contrast-enhanced vs non-contrast images). To address this issue, in this paper, we instead assume that a small representative dataset can be shared with the clients, with the style most commonly encountered, and thus have the clients learn an \emph{unpaired} domain mapping between the local and target domains using a CycleGAN. Our contribution to the state-of-the-art is two-fold:

% Closely related work to ST-FL is the research of Yang~\etal~\cite{yang2021federated}, who propose a semi-supervised federated learning framework for chest CT scan segmentation that leverages both labelled and unlabelled data at client nodes and is evaluated over multi-national data from China, Italy and Japan. This work shows the benefit of exploiting unannotated CT scan images in an FL setup for the task of image segmentation. In contrast to our work they do not address the mixed data issue and the fact that in realistic scenarios the CT images on each client node can vary massively in quality. In other recent related work Jeong~\etal~\cite{jeong2020federated} propose a semi-supervised FL framework that tackles the issue where private client data contains only partial or no labels. Data normalisation is tackled in~\cite{Kumar20} but they employ fixed transformations~\eg Lanczos interpolation, to standardise the hetrogenous client data, whereas we exploit the non-linear mappings possible through deep neural networks.

\begin{itemize}
    \item \textbf{Mixed CT image data quality \& the effect on FL}: Through experimentation with synthetic and semi-synthetic datasets of varying structural and stylistic features, we highlight the negative effect of differing quality images on client nodes on the accuracy of a federated U-Net \cite{Ronneberger15} for CT image segmentation.
    \item \textbf{Noise agnostic FL for varying noise patterns:}  We present \emph{ST-FL}, a federated learning framework that incorporates the denoising CycleGAN at each client node, standardising image quality per client and increasing the robustness of federated learning to mixed data quality observed in practice. For \emph{normalising} the image quality on client nodes with a CycleGAN \cite{zhu2017unpaired}, we propose two approaches. i) \emph{Universal CycleGAN}: only one denoiser is trained at the aggregator level and is then shared with the clients. ii) \emph{Client-specific CycleGAN}: multiple client-specific denoisers are trained at a client level. Experimental evaluation shows that ST-FL leads to higher quality segmentation models for chest CT scan images.
\end{itemize}

%%% METHODOLOGY
% \vspace{-3mm}
\section{METHODOLOGY}
\label{sec:methodology}
We used a number of publicly available COVID-19 segmentation datasets, which include segmentation masks generated by radiologists. We extracted a small amount of data to be our \emph{target style} and used the rest for training and testing. In addition to the already-existing noise patterns of the dataset (\eg discolouration, blurring, contrast), we further enhanced them with artificial noise (\eg contrast enhancement, contrast inversion, Gaussian noise, mixed noise \etc). We experimented with the Universal Cycle-GAN and Client-specific Cycle-GAN approaches and compared the thresholded segmentation results with respect to the FedAvg scheme and a Centralised model trained on style transferred datasets. The CycleGANs consist of a U-Net generator and a PatchGAN \cite{isola2018imagetoimage} discriminator. Prior to the actual federated training, we train both Universal and Client Specific CycleGANs for 100 epochs. For federated training, we concatenate original and style transferred images for each client segmentation UNet, ensuring that client models can learn salient information from each channel. This also ensures that the performance of the model is \textit{at least} comparable to FedAvg, because the model weights will be adapted to consider information only from the original input channel in the worst case.  For client datasets that serve as style targets, we concatenate 2 copies of the same image to input in the local segmentation model. We then train these models in a federated setting for 35 epochs. At the end of each training epoch, we aggregate their weights in a server model and broadcast them back to each client.

% \label{subsec:methodology}

% (Figure 2 metrics are averaged across number of trial runs and clients (tested on 3, 4, and 5 clients)

%%% EXPERIMENTS & RESULTS
\vspace{-1.5mm}
\section{EXPERIMENTS \& RESULTS}
\label{sec:results}

\subsection{Datasets}
In order to test the efficacy of our scheme, we perform experiments with 2 different types of client datasets:
\begin{itemize}
    \item \textbf{Synthetic Dataset}: We use the Coronacases \cite{jun2020covid} dataset of COVID-19 patient chest scans, both in its vanilla form and in an augmented form wherein each client dataset represents different noise patterns added to the dataset (inversion, Gaussian, contrast enhanced, mixed, etc). This scenario models a situation where client institutions may have chest scans with similar structural characteristics but differing style characteristics.
    \item \textbf{Semi-Synthetic Dataset}: We use the Coronacases and MedSeg\cite{COVID-19} dataset as client datasets and augment them with similar noise patterns as above to create additional client datasets. Compared to the Coronacases dataset, the MedSeg dataset was seen to have noisy labels and some structural and stylistic differences that can potentially hinder effective training.
\end{itemize}
% \begin{figure}[H]
%     \centering
%     \includegraphics[scale=0.65]{LaTeX/images/synthetic_dataset_example.PNG}
%     \caption{Sample Images from Synthetic Datasets: (Left to Right: Vanilla Coronacases, Mixed Noise Coronacases, Noisy Coronacases, Inversion Coronacases, Contrast Enhanced Coronacases - Style Target)}
%     \label{fig:synthetic_dataset}
% \end{figure}
% \begin{figure}[H]
%     \centering
%     \includegraphics[scale=0.66]{LaTeX/images/semi-synthetic_dataset_example.PNG}
%     \caption{Sample Images from Semi-Synthetic Datasets: (Left to Right: Vanilla Medseg - Style Target, Vanilla Coronacases, Noisy Coronacases, Inversion MedSeg, Mixed Noise Coronacases)}
%     \label{fig:semi_synthetic_dataset}
% \end{figure}
In this paper, we consider the scenario where client datasets are of similar size. Because the Coronacases and MedSeg datasets are of different sizes (30 and 100 images respectively), we fix $|\mathcal{D}^k| = 30$ for all clients. For both approaches, we add random warping to all client images not only for data augmentation, but also to add some variability in different client datasets. This also ensures that the CycleGAN is able to learn unpaired mappings between the original and target styles and becomes agnostic to any structural similarity between images. After warping, we keep aside $20\%$ of each client dataset $\mathcal{D}^k_{val}$ as a validation set and calculate the average Dice and IOU score on the union of all client validation sets $\mathcal{D}_\textrm{val} = \cup_{k=1}^N$ $\mathcal{D}^k_{val}$. To understand which noise patterns the CycleGAN based approaches perform well on, we calculate the average performance improvement on the union of training and validation sets for each client (Figure \ref{fig:bar_plot}).

\subsection{Results}
Table \ref{tab:metric_table} shows metric scores of the different schemes tested across clients and dataset types. For all dataset types and number of clients, we observe that the client specific CycleGAN preprocessing scheme outperforms its federated learning counterparts. For synthetic datasets with high degree of structural similarity, the centralised training scheme can be viewed as an upper bound on the segmentation performance relative to all other federated learning schemes, with the client specific CycleGAN scheme coming closest to this bound. For semi-synthetic datasets, on the other hand, we observe no discernable pattern in segmentation performance of centralised training, with the client specific CycleGAN scheme outperforming it in all cases. Intuitively, this is because the centralised model is being trained on datasets of varying structural similarities and noisy labels, making it harder for weights to generalise across datasets. The client specifc CycleGAN becomes more robust to this, leading to improved performance across all noise pattens tested. We also see that universal CycleGAN preprocessing offers lower and more inconsistent performance gains on average compared to the client specifc CycleGAN scheme. This is because the style transferred output from the Universal CycleGAN was often of poor quality, especially in situations involving large numbers of clients, as the model is unable to adapt to differing client distributions.
% \begin{figure}[H]
%     \centering
%     \includegraphics[scale=0.7]{LaTeX/images/style_transfer_comparison.PNG}\\
%     \includegraphics[scale=0.7]{LaTeX/images/style_transfer_comparison_3.PNG}
%     \caption{Comparison Between Style Transferred Images for Cycle-GAN Based Pre-Processors}
%     \label{fig:style_transfer_comparisons}
% \end{figure}
% This is further reinforced in Figure \ref{fig:bar_plot}, where the left bar plot indicates that 
% Double horizontal break + (federated and non federated section) + reinforce centralised is the upper bound
% \usepackage{multirow}
% \usepackage{graphicx}
% \begin{table}[H]
% \centering
% \resizebox{\textwidth}{!}{%
% \newpage
{\small\tabcolsep=3pt
\begin{longtable}{c|c|c|ccc|c}
\hline
\multirow{3}{*}{\textbf{Number of Clients}} &
  \multirow{3}{*}{\textbf{Metric}} &
  \multirow{3}{*}{\textbf{Dataset Type}} &
  \multicolumn{3}{c|}{\multirow{2}{*}{\textbf{Federated Training}}} &
  \multirow{3}{*}{\textbf{Centralised Training}} \\
                   &                       &                & \multicolumn{3}{c|}{}                                               &  \\ \cline{4-6}
 &
   &
   &
  Vanilla FedAvg &
  \begin{tabular}[c]{@{}c@{}}Universal \\ CycleGAN\end{tabular} &
  \begin{tabular}[c]{@{}c@{}}Client Specific \\ CycleGAN\end{tabular} &
   \\ \hline
\multirow{4}{*}{3} & \multirow{2}{*}{Dice} & Synthetic      & 0.414 $\pm$ 0.012  & 0.505 $\pm$ 0.027 & \textbf{0.533} $\pm$ \textbf{0.041} & 0.560 $\pm$ 0.023 \\
                   &                       & Semi-Synthetic & 0.497 $\pm$ 0.009  & 0.520 $\pm$ 0.012 & \textbf{0.539} $\pm$ \textbf{0.012} & 0.494 $\pm$ 0.009 \\ \cline{2-7} 
                   & \multirow{2}{*}{IOU}  & Synthetic      & 0.274 $\pm$ 0.011 & 0.329 $\pm$ 0.017 & \textbf{0.336} $\pm$ \textbf{0.019} & 0.347 $\pm$ 0.016  \\
                   &                       & Semi-Synthetic & 0.337 $\pm$ 0.014 & 0.355 $\pm$ 0.016 &\textbf{ 0.366} $\pm$ \textbf{0.006} & 0.338 $\pm$ 0.010 \\ \hline
\multirow{4}{*}{4} & \multirow{2}{*}{Dice} & Synthetic      & 0.430 $\pm$ 0.008 & 0.493 $\pm$ 0.026 &\textbf{ 0.514} $\pm$ \textbf{0.037} & 0.528 $\pm$ 0.017  \\
                   &                       & Semi-Synthetic & 0.462 $\pm$ 0.014 & 0.483 $\pm$ 0.021 &\textbf{ 0.489} $\pm$ \textbf{0.014} & 0.450 $\pm$ 0.013  \\ \cline{2-7} 
                   & \multirow{2}{*}{IOU}  & Synthetic      & 0.287 $\pm$ 0.007 & 0.296 $\pm$ 0.013 &\textbf{ 0.332 }$\pm$ \textbf{0.017} & 0.329 $\pm$ 0.007  \\
                   &                       & Semi-Synthetic & 0.306 $\pm$ 0.007 & 0.319 $\pm$ 0.006 & \textbf{0.321} $\pm$ \textbf{0.004} & 0.305 $\pm$ 0.008  \\ \hline
\multirow{4}{*}{5} & \multirow{2}{*}{Dice} & Synthetic      & 0.378 $\pm$ 0.012 & 0.460 $\pm$ 0.025 &\textbf{ 0.465} $\pm$ \textbf{0.006} & 0.490 $\pm$ 0.006  \\
                   &                       & Semi-Synthetic &  0.390 $\pm$ 0.018 &  0.420 $\pm$ 0.017 & \textbf{0.462} $\pm$ \textbf{0.017} & 0.392 $\pm$ 0.018 \\ \cline{2-7} 
                   & \multirow{2}{*}{IOU}  & Synthetic       & 0.255 $\pm$ 0.008 & 0.240 $\pm$ 0.009 & \textbf{0.282} $\pm$ \textbf{0.013} & 0.310 $\pm$ 0.012  \\
                   &                       & Semi-Synthetic & 0.265 $\pm$ 0.007 & 0.281 $\pm$ 0.012 & \textbf{0.301} $\pm$ \textbf{0.010} & 0.261 $\pm$ 0.010  \\ \hline

\caption{Performance Metrics for the Methods Tested for Differing Numbers of Clients. Note that we report the best metric averaged over 5 trials and its associated 95$\%$ confidence interval.}
\label{tab:metric_table}
\end{longtable}%
}

\begin{figure}[H]
    \centering
    \includegraphics[scale=0.51]{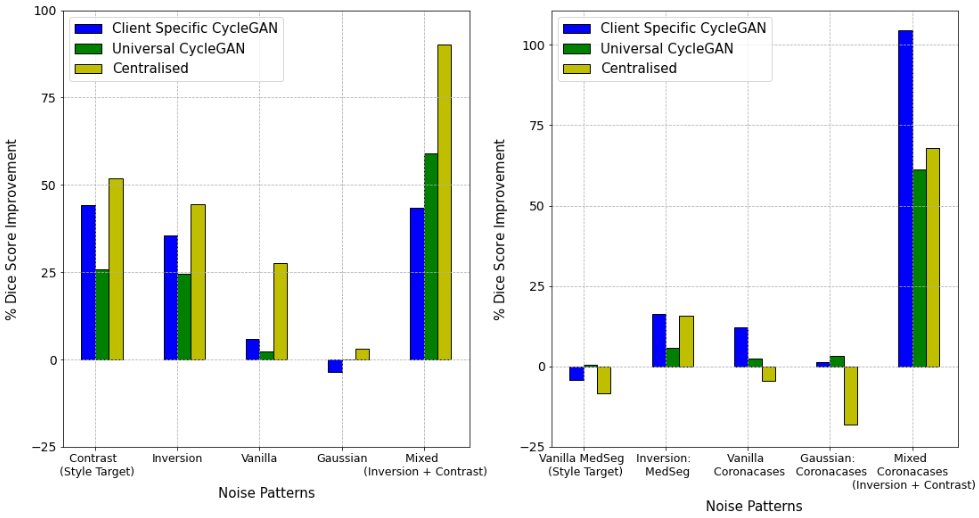}
    \caption{Best Case $\%$ Noise-Specific Performance Improvement of the Different Techniques Tested Relative to Vanilla FedAvg (Left: Synthetic Datasets, Right: Semi-Synthetic Datasets)}
    \label{fig:bar_plot}
\end{figure}

% \end{table}
% \varun{
% (Showing average $\%$ performance improvement of CS-GAN, Uni-GAN, and Centralised over FedAvg for synthetic and semi-synthetic datasets with 95$\%$ confidence interval, segmentation visualisations, table of best DICE / IOU Scores with 95$\%$ confidence interval, validation curves for the above methodologies averaged over number of clients.)\tony{TODO}}\\
Figure \ref{fig:bar_plot} shows a bar plot of the performance gains of all the scheme tested relative to the vanilla FedAvg scheme. Here, we averaged the $\%$ performance improvement of each noise pattern across differing numbers of clients tested, over $5$ runs. We note that there is a discrepancy in the performance of CycleGAN related schemes over the noise patterns tested. Specifically, for both the semi synthetic and synthetic dataset, applying style transfer preprocessing on images corrupted by Gaussian noise does not produce meaningful improvement in dice scores. This is likely due to information loss in images that is not necessarily corrected by unpaired style transfer. Conversely, we see significant gains in performance in inversion and mixed noise patterns for both dataset types, though these gains are larger for the synthetic dataset case. Intuitively, this is because averaging weights of models trained on structurally dissimilar client datasets likely limits the benefits of style transfer, which can only correct for noise distribution shifts and not structural shifts.
\vspace{-3.5mm}
\section{NOVEL WORK \& CONCLUSIONS}
This paper presents a novel preprocessing method for performing federated learning in a noise agnostic manner, with a focus on segmentation of lesions in COVID 19 patient chest scans. Medical datasets in a federated learning setup tend to have variations in contrast, noise, brightness and detail, motivating the need for a common normalisation scheme which renders federated systems agnostic to noise. We explored the idea of using style transfer based pre-processing on client datasets in 2 scenarios: a) varying noise patterns but common structure, and b) varying noise patterns and varying structure. Our work suggests that style transfer pre-processing leads to higher dice scores in downstream segmentation tasks on average in both cases. We characterised the performance of our method on some common noise patterns in medical datasets and found disparities in performance, with some noise patterns showing much greater improvement in segmentation performance than others. Future work could focus on exploration of this technique in settings where client datasets are unbalanced and / or are of unequal size and further characterise its noise-specific performance.
\bibliography{main} % bibliography data in report.bib
\bibliographystyle{spiebib} % makes bibtex use spiebib.bst

\end{document}